\documentclass[aps,prl,superscriptaddress,amsfonts,amssymb,amsmath,nofootinbib,twocolumn,floatfix]{revtex4}
 \usepackage{color,graphicx}
 \usepackage[utf8]{inputenc}
 \usepackage[T2A]{fontenc}
 \usepackage[russian,english]{babel}
 \definecolor{darkblue}{rgb}{0,0,0.7}
\definecolor{darkred}{rgb}{0.7,0,0}
\definecolor{darkgreen}{rgb}{0,0.4,0}
\usepackage[unicode, colorlinks, citecolor=darkblue, linkcolor=darkred, urlcolor=blue]{hyperref}

\allowdisplaybreaks
\begin{document}

\author{Sergey P. Vyatchanin}

\affiliation{Faculty of Physics, M.V. Lomonosov Moscow State University, Leninskie Gory, Moscow 119991, Russia}
\affiliation{Quantum Technology Centre, M.V. Lomonosov Moscow State University, Leninskie Gory, Moscow 119991, Russia}

\author{Andrey B. Matsko}

\affiliation{Jet Propulsion Laboratory, California Institute of Technology, 4800 Oak Grove Drive, Pasadena, California 91109-8099, USA}

\date{\today}
	
\title{Broadband Back Action Cancellation in a Dichromatic Variational Measurement}

\begin{abstract}
Quantum back action imposes fundamental sensitivity limits to the majority of quantum measurements. The effect results from the unavoidable contamination of the measured parameter with the quantum noise of a meter. Back action evading measurements take advantage of the quantum correlations introduced by the system under study to the meter and allow overcoming the fundamental limitations. The measurements are frequently restricted in their bandwidth due to a finite response time of the system components. Here we show that probing a mechanical oscillator with a dichromatic field with frequencies separated by the oscillator frequency enables independent detection and complete subtraction of the measurement noise associated with the quantum back action. 
\end{abstract}

\maketitle

Detection of a classical force acting on a free mass or a mechanical oscillator using an optical meter is one of the major problems of modern optomechanics. It is related to precision measurements of gravitational waves \cite{aLIGO2013,aLIGO2015}, gravity \cite{rademacher20aot}, acceleration \cite{huang20lpr}, torque \cite{WuPRX2014, KimNC2016, AhnNT2020}, and  magnetic field \cite{ForstnerPRL2012, LiOptica2018}. The problem is also linked to the frequency conversion between microwaves and light \cite{LambertAQT2019, LaukQST2020} and to the testing the limits of quantum mechanics \cite{BassiRMP2013, ArndtNP2014}. The fundamental sensitivity of the measurement is restricted by so called standard quantum limit (SQL) \cite{Braginsky68,BrKh92}, unless a special measurement procedure is utilized. In this Letter we introduce a new technique allowing beating the SQL in a broad frequency range.

SQL was studied in many physical systems ranging from kilometer-sized gravitational wave detectors \cite{02a1KiLeMaThVyPRD} to microcavities \cite{Kippenberg08,DobrindtPRL2010}. For instance, let us consider the dispersive optomechanical interaction \cite{AspelmeyerRMP2014,Marquardt09} realized in an optical cavity where position of a mechanical body (moving mirror) modifies the cavity eigenfrequency. A force proportional to the optical power or to the number of optical quanta inside the optical cavity occurs due to the light pressure. In the case of a coordinate measurement the SQL results from the noise introduced by the pressure force. 
Since the information about the coordinate is recorded in the phase of the light reflected from the mechanical system, the signal is masked by the phase noise of the readout. The radiation pressure noise also masking the signal is proportional to the amplitude fluctuations of the readout. Phase and amplitude fluctuations of the same monochromatic wave do not commute and, as the result, cannot be simultaneously reduced to zero. Quantum theory does not allow accurate simultaneous measurements of the amplitude and phase fluctuations of the readout. This leads to the SQL of the coordinate measurements of a force. 

The SQL does not pose a general fundamental limit for the classical force detection. It can be avoided by applying a variational measurement technique \cite{93a1VyMaJETP, 95a1VyZuPLA, 02a1KiLeMaThVyPRD}. Preparation of the probe light in an quantum squeezed state leads to the optimum measurement sensitivity better than SQL \cite{LigoNatPh11, LigoNatPhot13, TsePRL19, AsernesePRL19, YapNatPhot20, YuNature20, CripeNat19}. The limit also can be surpassed  with coherent quantum noise cancellation technique \cite{TsangPRL2010, PolzikAdPh2014, MollerNature2017} and utilization of optical rigidity \cite{99a1BrKhPLA, 01a1KhPLA}.

Optomechanical systems having several degrees of freedom in the optical probe, mechanical system or both, feature more complex interactions including radiation puling (negative radiation pressure) \cite{Povinelli05ol,maslov13pra}, ponderomotive force proportional to a generalized optical quadrature \cite{93a1VyMaJETP,96a1VyMaJETP,matsko97apb,02a1KiLeMaThVyPRD}, as well as mechanical velocity (rather than coordinate) dependent interaction \cite{90BrKhPLA,00a1BrGoKhThPRD}. 
Polychromatic probe can facilitate force measurements with better than SQL sensitivity \cite{80a1BrThVo, 81a1BrVoKh}. This technique was adopted for a cavity with a movable test mass probed with the light modulated at frequency twice of the mechanical frequency \cite{Clerk08, Clerk2015, Pirkkalainen2015, 18a1VyMaJOSA}. Usage of two tone local oscillator for analysis of the signal produced in the cavity interrogated with a monochromatic light wave results in back action evading measurement with sensitivity limited by thermal noise \cite{Buchmann2016}. 

We here propose to merge the variational measurement and dichromatic interrogation techniques to improve the detection of the force. We have realized that illuminating a mechanical oscillator with a dichromatic wave and properly measuring the signal using ideas of the variational approach results in decoupling the measured signal from the back action noise. As a result, the back action noise can be detected {\em separately } from the signal and subtracted from the measured signal while postprocessing the measurement results. Such a measurement allows detection of the mechanical force with sensitivity better than SQL. 

The technique described in what follows has several novel features. A standard variational measurement has a narrow bandwidth shrinking with the measurement sensitivity improvement. One cannot overcome SQL in a broad frequency band without usage of a special signal post processing technique. Moreover, the variation measurement is applicable only if the signal force acts on a free test mass, it cannot be used for detection of a resonant force acting on a mechanical oscillator.  The technique described in this Letter resolves both issues. We show that the measurement can be broadband and also that it can be utilized for detection of a resonant force. 

The main message of our study is that in the case of a polychromatic probe field the quantum back action can depend on the quantum fluctuations that commute with the measured observable of the probe field containing the signal and, hence, can be measured independently and subtracted from the measurement result. Such a measurement is not feasible in the ordinary case when the back action results from the amplitude fluctuations of the monochromatic probe wave while the signal is recorded in the phase of the probe wave, as the amplitude and phase of the same wave cannot be measured simultaneously in the same frequency band. In what follows, we illustrate our finding considering the case of a displacement transducer comprising a mechanical oscillator interrogated with either monochromatic or polychromatic light and using a homodyne detection scheme with either monochromatic or polychromatic local oscillator. We start from a brief review of the standard variational approach and then introduce the dichromatic probe technique.

Let us consider a scheme in which a monochromatic optical probe wave with carrier frequency $\omega_0$ is reflected from a movable mirror being a part of the mechanical oscillator of frequency $\omega_M$ with mass $m$. The information about a force of interest $F_S(t)$ is obtained by the detection of the optimal quadrature amplitude of the reflected light. We present amplitude of the probe wave as $\mathcal A  = \left(A + \hat a\right) \exp (-i\omega_0 t)$ and normalize it so that $P_0=\hslash \omega_0 |A|^2$ becomes the averaged power of the probe wave \cite{02a1KiLeMaThVyPRD}. Considering coherent probe we can write $\left[\hat a(t), \hat a^\dag(t')\right] =   \delta(t-t')$ and $\left\langle\hat a(t) \hat a^\dag(t')\right\rangle = \delta(t-t')$, where $\langle \dots \rangle$ stands for ensemble average. Introducing Fourier transform for the annihilation operator as $ \hat a (t) = \int_{-\infty}^\infty a(\omega) \, \exp(-i\omega t)\, d\omega/(2\pi)$, we get corresponding relationships for the Fourier amplitudes: $\left[ a(\omega),  a^\dag(\omega')\right] = 2\pi\,\delta(\omega -\omega')$ and $\left\langle a(\omega)\,  a^\dag(\omega')\right\rangle = 2\pi\, \delta(\omega -\omega')$.

We present the coordinate of the movable mirror as $\hat x_M = X+ x_0 \hat x$, where $X$ is the shift of the mirror due to the radiation pressure effects and $\hat x$ stands for the dimensionless fluctuation of the mirror position, the normalization constant is $x_0 = (\hbar/(2m \omega_M))^{1/2}$. The radiation pressure due to the monochromatic probe introduces constant shift of the mirror position $X=2P_0/mc\omega_M^2$. The fluctuations of the mirror coordinate are described by
\begin{equation} \label{x1}
    \hat {\ddot  x}+2 \gamma_M \hat {\dot x} +\omega_M^2 \hat x=\frac{2\hslash k}{mx_0}
     (A\hat a^\dag+A^*\hat a)+2\omega_M (f_s+f_{fl})
\end{equation}
where $\gamma_M$ is the mechanical relaxation rate, $k= \omega_0/c$ is the optical wave number, $f_s=F_S/(2\hbar \omega_M m)^{1/2}$ is the normalized signal force, $f_{fl}(t)=\int_0^\infty (2\gamma_M \omega/\omega_M)^{1/2} (e_{fl}(\omega) \exp(-i\omega t)+h.c.) (d\omega/2\pi)$ is the fluctuation force associated with coupling of the mechanical oscillator to the thermal bath.  The Fourier amplitudes of the thermal operators obey to the relationships  $\left[ e_{fl}(\omega),  e_{fl}^\dag(\omega')\right] = 2\pi\,\delta(\omega -\omega')$ and $\left\langle e_{fl}^\dag(\omega)\,  e_{fl}(\omega')\right\rangle = 2\pi\, n_{T}\, (\omega) \delta(\omega -\omega')$, $n_T=(\exp(\hbar \omega/k_BT)-1)^{-1}$ is the thermal quanta occupation number, $T$ is the reservoir temperature, and $k_B$ is the Boltzmann constant. Fourier amplitudes of the coordinate and the signal force are introduced as $\hat x=\int_{-\infty}^\infty x(\omega) \, \exp(-i\omega t)\, d\omega/(2\pi)$, $f_s(t)=\int_{-\infty}^\infty f_s(\omega) \, \exp(-i\omega t)\, d\omega/(2\pi)$.

For the reflected wave we obtain the following expression in the linear approximation by $\hat x$:
\begin{equation} \label{bout1a}
\hat b = \hat a + 2ikx_0A \hat x,
\end{equation}
where $\hat b$ ($\hat a$) is output (input) amplitude. The time independent phase shift due to the constant radiation pressure is neglected since it can be compensated.  
The generalized homodyne scheme measures combination $b_\psi = b_a\cos\psi + b_\phi \sin\psi$ of amplitude ($b_a$) and phase ($b_\phi$) quadratures of the reflected wave described with Fourier amplitudes $b_a=\left(b(\omega)+b^\dag(-\omega)\right)/\sqrt 2,\ b_\phi=\left(b(\omega)-b^\dag(-\omega)\right)/i\sqrt 2 $.
Solving Eqs.(\ref{x1}) and (\ref{bout1}) and assuming $A=A^*$, we derive for the output field amplitude
\begin{eqnarray} \label{bout1}
        b(\omega)=a(\omega)+\frac{i{\cal K}}{\sqrt{2}Z}a_a+\frac{i\sqrt{{\cal K} \omega_M}}{Z}\big(f_s(\omega)+f_{fl}(\omega)\big), 
\end{eqnarray}
and associated quadratures
\begin{align}
    \label{ba}
    b_a &= a_a,\\
    \label{bphi}
    b_\phi &= a_\phi + \frac{{\cal K}}{Z}\,  a_a+ \frac{\sqrt{\mathcal K \omega_M}}{Z}\,\big(f_{s\, a}+f_{fl\, a}\big), 
\end{align}
where ${\cal K}=8\hslash k^2A^2/m$, $Z=\omega_M^2-\omega^2-2i\gamma_M\omega$, and where $f_{s\, a}$ and $f_{fl\, a}$ are the amplitude quadratures of the forces. Equations (\ref{x1}-\ref{bphi}) are general. For example, for the case of the free mass approximation we should assume $\omega \gg \omega_M$. 

It is not possible to measure $b_a$ and $b_\phi$ simultaneously since they do not commute. Let us measure phase quadrature $b_\phi$ assuming that the (normalized) narrow-band force of interest $f_S(t)=f_{S0}\cos{\omega_{f0} t}$ acts on the movable mirror when $\tau/2 \geq t \geq -\tau/2$. The signal is detectable when
\begin{align}
 \label{fs2}
 f_{S0} & \ge  \sqrt{S_n (\omega_{f0}) \Delta \omega/2\pi}, \\
 S_n (\omega) & =\frac{ 2\gamma_M \omega(2n_T +1)}{\omega_M} 
   + \frac{ |Z|^2}{2\mathcal K \omega_M} +\frac{\mathcal K}{2\omega_M},
\end{align}
where $S_n$ is single-sided power spectral density of the fluctuations and $\Delta \omega \simeq 2 \pi/\tau$ is the measurement bandwidth. The minimal noise of the force measurement is achieved when $\mathcal K= |Z(\omega_{f0})|$:
\begin{align} \label{sql1}
 S_n (\omega_{f0}) &= \frac{2\gamma_M \omega_{f0} (2n_T +1)}{\omega_M}+ \frac{ |Z(\omega_{f0})|}{ \omega_M}.
\end{align}
At $\gamma_M=0$ the spectral density \eqref{sql1} corresponds to the SQL of the force.

The term proportional to $\mathcal K$ in \eqref{bphi} stands for the quantum back action. It can be cancelled around the fixed frequency $\omega_{f0}$, if one measures the quadrature $b_\psi= b_a\cos \psi + b_\phi\sin\psi$ with optimally selected  angle $\psi$ which fulfills condition $\cos \psi + Re[\mathcal K /Z(\omega_{f0})]\sin \psi=0$. In this case
\begin{align}
 S_n (\omega_{f0}) &= \frac{2\gamma_M \omega_{f0}(2n_T +1)}{\omega_M}  + \frac{ |Z(\omega_{f0})|^2}{2\mathcal K \omega_M}.
\end{align}
$S_n(\omega_{f0})$ monotonically decreases with increasing power if $\gamma_M=0$. This is true in a narrow frequency band $\delta \omega \simeq |Z(\omega_{f0})|^2/(2 \omega_{f0} \mathcal K)$. To achieve a broadband detection a frequency dependent $\psi$ should be utilized \cite{02a1KiLeMaThVyPRD}. Parameter $Z(\omega_M)$ is imaginary for the resonant force and back action compensation is not possible. The thermal noise limits the sensitivity unless the signal frequency $\omega_{f0}$ is far from the resonance frequency: $|\omega_{f0} -\omega_M| \gg \Delta \omega \gg \gamma_M$ \cite{93a1VyMaJETP}. 
 
\begin{figure}
    \includegraphics[width=0.18\textwidth]{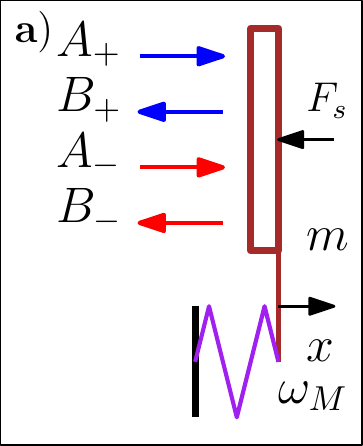}\
    \includegraphics[width=0.275\textwidth]{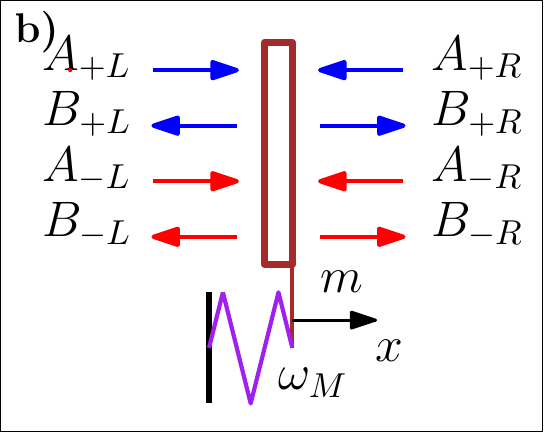}
    \caption{{\em a)} A toy model explaining a dichromatic variational measurement of a test mass with two probes having frequencies $\omega_+,\ \omega_-$. The test mass is a suspended perfectly reflecting mirror characterized with frequency $\omega_M$ and mass $m$. Frequencies $\omega_+,\ \omega_-$ relate to mechanical frequency as $\omega_+-\omega_-= \omega_M $. A classical feedback force has to be applied to the mirror to avoid resonant excitation of the mechanical oscillation.
    {\em b)} Four monochromatic probe fields are incident on the same test mass from the opposite sides. In this scheme the resonant ponderomotive mechanical excitation of the mirror is compensated. }\label{NoCav}
   \end{figure}

The standard variational approach is not optimal for the resonant force detection. To surpass the SQL of the force measurement we suggest to utilize a dichromatic probe field interrogating a moving mirror which is a part of a mechanical oscillator.  Let us consider first a toy model in which the displacement of the mechanical oscillator is measured via registering quadratures of the two reflected light waves with carrier frequencies $\omega_\pm=\omega_0 \pm \omega_M/2$  (see Fig.~\ref{NoCav}a) by using corresponding balanced homodyne detectors. A classical resonant force $F_{comp}$ is applied to the mirror to suppress its oscillation resulting from the ponderomotive resonant excitation.

We denote the amplitudes of the incident and reflected waves as $(A_{\pm}+\hat a_{\pm}) \exp (-i \omega_\pm t)$ and $(B_{\pm}+\hat b_{\pm}) \exp (-i \omega_\pm t)$ respectively, and write the mirror position in the form $\hat x_M = x_0 ([D+\hat d] e^{-i\omega_M t}+ [D^*+ \hat d^\dag]e^{i\omega_M t})$, where $\hat a_\pm$, $\hat b_\pm$, $\hat d$ stand for the annihilation operators of the fields as well as the mirror position. We also assume that there is a known classical resonant force $F_{comp}e^{-i\omega_M t}+c.c.$ acting on the mirror. The force is applied to suppress resonant oscillations of the probe mass.

The expectation values obey two expressions obtained using Eqs.~(\ref{x1}) and (\ref{bout1a}), where the monochromatic field is replaced with the dichromatic one:
\begin{align}
& B_\pm=A_\pm \left ( 1 \mp \frac{4k^2x_0^2|A_\mp|^2}{\gamma_M}   \right ),\\
& \gamma_M D= 2ikx_0A_+A_-^*+F_{comp}.
\end{align}
These expressions show that the measurement scheme supports excitation of the mechanical oscillation if $F_{comp}=0$. For identical probe harmonics, $A_\pm = A$, this effect can be neglected, if $1 \gg {\mathcal K}/4\gamma_M \omega_M$. However, this condition prevents us from beating the SQL. On the other hand, if we assume that the classical feedback force $F_{comp}$ is selected so that $2ikx_0A_+A_-^*+F_{comp}=0$ and $D=0$, we can show below that in this case the SQL can be beaten.

For the Fourier amplitudes of the fluctuations we write input-output relations
 \begin{align}
  \label{b+1}
  b_+(\Omega) &= a_+(\Omega) + 2ikx_0 A \,d(\Omega),\\
  \label{b-1}
  b_-(\Omega) &= a_-(\Omega) + 2ikx_0 A d^\dag(-\Omega),\\
  \label{dOmega1}
 d(\Omega) &=  \displaystyle \frac{2i\, kx_0A}{\gamma_M-i\Omega } 
	 \left(a_-^\dag(-\Omega) +  a_+(\Omega)\right) + \\ \nonumber & \displaystyle \frac{ i[ f_s(\Omega)+f_{fl}(\Omega)]}{\gamma_M-i\Omega}, 
 \end{align}
where $a_\pm(\Omega) \equiv a(\Omega\pm\omega_M/2)$ and $a_\pm^\dag(-\Omega) \equiv a^\dag(-\Omega\pm\omega_M/2)$ and the same for $b_\pm(\Omega)$. 

Let us introduce quadrature amplitudes of the output waves as $b_{\pm a}=(b_\pm(\Omega)+b_\pm(-\Omega))/ \sqrt 2$. Linear combinations of the commuting amplitude quadratures $\beta_{a\pm}=(b_{+ a} \pm b_{-a})/ \sqrt 2$  {\em can be measured simultaneously}. We obtain for them
\begin{align}
    \label{betaplus}
        \beta_{a+} &= \alpha_{a+}, \\
    \label{betaminus}
        \beta_{a-} &= \alpha_{a-}- \frac{{\mathcal K}}{2\omega_M(\gamma_M-i\Omega)} \alpha_{a+}- \\
        &\qquad \nonumber \frac{\sqrt{\mathcal K}}{\sqrt {2 \omega_M}(\gamma_M-i\Omega)} (f_{S, a}+f_{fl, a}).
\end{align}
We notice that to perform the optimal measurement one has to detect and record simultaneously $\beta_{a+}$ and $\beta_{a-}$ and then evaluate their linear combination 
\begin{align} \nonumber
   B_\beta &=\beta_{a-}+ \frac{{\mathcal K}}{2\omega_M(\gamma_M-i\Omega)} \beta_{a+}  \\
   & =\alpha_{a-}- \frac{\sqrt{\mathcal K}}{\sqrt {2\omega_M}(\gamma_M-i\Omega)} (f_{S, a}+f_{fl, a}) 
\end{align}
In such a measurement the back action is completely excluded and the measurement sensitivity is defined by the spectral density
\begin{align}
 S_n (\Omega) &\ge 2\gamma_M (2n_T +1)  + \frac{ 4(\gamma_M^2+\Omega^2) \omega_M}{\mathcal K }.
\end{align}
Remarkably, the back action is evaded in a broad frequency range, unlike the case of the conventional variational measurements. The thermal noise term can be filtered out at small $\gamma_M$ and the SQL can be beaten if $|\Omega| \gg \gamma_M$. This result can be repeated with the phase quadratures.

For the toy measurement scheme above the sensitivity is formally limited because of the resonant growth of the amplitude $D$ that prevents the increase of the sensitivity with an optical power increase, unless an optimal classical resonant force is applied to the mirror to compensate for the unlimited classical amplitude growth. An alternative  quantum mechanochal approach involves two dicromatic waves incident on the opposite surfaces of the mirror mass (see Fig.~\ref{NoCav}b). It allows to perform the measurement without the classical feedback force. 

We select the dichromatic probes characterized by a real amplitude $A$ and frequencies  $\omega_0 \pm \omega_M/2$, the same as previously. For symmetry reasons we conclude that $D=0$ in this case.  Two more harmonics centered at frequencies $\omega_0 \pm 3\omega_M/2$ have to be taken into account in addition to the main probe fields. The fluctuations in other frequency bands can be neglected. Using Eq.~(\ref{x1}) we derive an expression for the spectral amplitude of the mirror position
\begin{align} \label{dd}
    d(\Omega)&=\frac{i(f_S(\Omega)+f_{fl}(\Omega))}{\gamma_M-i\Omega}+\frac{2ikx_0A}{\gamma_M-i\Omega}  \times \\ 
    \nonumber &\times \sum \limits_{l,n=1,2} (-1)^{l-1} \left [a_{l,+n}\left (\Omega\right ) + a_{l,-n}^\dag \left (-\Omega \right )\right ]. 
\end{align}
where $l=1$(2) stands for the left (right) probe waves, respectively; $a_{l,\pm n} =a_l(\Omega \pm (2n-1) \omega_M/2  )$. Expressions for the spectral amplitudes of the output field fluctuations 
\begin{align} \label{blp}
      & b_{l,+n} \left ( \Omega \right ) = a_{l,+n} \left (\Omega \right )+(-1)^{l-1} 2ikx_0 A d(\Omega), \\ \label{blm}
      & b_{l,-n} \left (\Omega \right ) = a_{l,-n} \left (\Omega \right )+(-1)^{l-1} 2ikx_0 A d^\dag (-\Omega).
\end{align}
along with Eq.~(\ref{dd}) lead to
\begin{align}
  & \alpha_{a\pm,l,n}= \\ \nonumber & \frac{1}{2}  \left ( a_{l,+n} \left (\Omega \right ) + a_{l,+n}^\dag \left (-\Omega \right ) \pm  a_{l,-n} \left (\Omega \right ) \pm a_{l,-n}^\dag \left (-\Omega \right )\right ), \\ \label{amp}
  & \beta_{a+,l,n}=\alpha_{a+,l,n},
  \\ \nonumber
  & \beta_{a-,l,n}=\alpha_{a-,l,n}\\ \nonumber 
  &\quad+(-1)^{l} \frac{{\mathcal K}}{2\omega_M(\gamma_M-i\Omega)} \sum \limits_{l1,n1=1,2} (-1)^{l1-1} \alpha_{a+,l1,n1}  \\  \label{pha}
  & \quad + (-1)^{l} \frac{\sqrt{\mathcal K}}{\sqrt {2\omega_M }(\gamma_M-i\Omega)} (f_{S,a}+f_{fl,a}).
\end{align}
Since the quadrature amplitudes  $\beta_{a+,l,n}$ and $\beta_{a-,l,n}$ commute, they can be measured independently. The signal can be found from a linear combination of these measurements
\begin{align} 
& B= \displaystyle \frac 1 2 \sum \limits_{l,n=1,2} (-1)^l \beta_{a-,l,n}= \\ \nonumber & \displaystyle \frac 1 2 \sum \limits_{l,n=1,2} (-1)^l \alpha_{a-,l,n} + \displaystyle \frac{\sqrt{2 \mathcal K}}{\sqrt {\omega_M}(\gamma_M-i\Omega)} (f_{S,a}+f_{fl,a})
\end{align}
that does not depend on the back action. We can measure four spectral components that reduces the error of the measurement by the factor of four, when compared with the single channel toy model 
\begin{align} \label{compen}
 S_n (\Omega) &\ge 2\gamma_M (2n_T +1)  + \frac{(\gamma_M^2+\Omega^2) \omega_M}{\mathcal K }.
\end{align}
This is an important result, since it shows that taking multiple frequency bands into account we can improve the fundamental sensitivity of the the measurement. 

To conclude, equation (\ref{compen}) is the main result of our study. We have shown that using a dichromatic variational measurement is advantageous for the detection of a classical force acting on a mass of a linear mechanical oscillator.  The peculiarity of the proposed technique is that the resonant dichromatic probe field enables the transfer of information from the quadrature of a mechanical oscillator to the commuting quadratures of the dichromatic optical field. One can measure the optical quadratures separately and then subtract out the back action while processing the measurement results. In our understanding, this is a new kind of measurement technique since usually the back action cannot be measured separate from the signal. In a standard scheme with a monochromatic pump the information about the positional coordinate is transferred to the phase of the probe wave. In this case, we have only one quantum output and the back action cannot be measured separate from the signal. Variational measurement allows the subtraction of the back action wthin a limited bandwidth. The strategy described in this Letter allows beating the SQL in a broad frequency band. Moreover, involving several spectral harmonics enables simultaneous independent measurements in different frequency bands that result in further sensitivity improvements. 

\acknowledgments
  Authors are grateful to Daniel Sigg for very valuable discussion and remarks.
  The research of SPV has been supported by the Russian Foundation for Basic Research (Grant No. 19-29-11003), the Interdisciplinary Scientific and Educational School of Moscow University ``Fundamental and Applied Space Research'' and from the TAPIR GIFT MSU Support of the California Institute of Technology. The research performed by ABM was carried out at the Jet Propulsion Laboratory, California Institute of Technology, under a contract with the National Aeronautics and Space Administration (80NM0018D0004). This document has LIGO number P2100141.


\end{document}